\documentclass[twocolumn,aps,showpacs,showkeys,groupedaddress,amsmath,amssymb,epsf]{revtex4}
\usepackage{graphicx}
\usepackage{dcolumn}
\usepackage{bm}

\def\R{{\mathbb R}}
\def\ep{\epsilon~}
\def\ro{R\"ossler~}
\def\ie{i.e.~}
\def\eg{e. g.~}

\def\etc{ etc.~}

\def\ep{\epsilon}
\def\eqn{\end{equation}\noindent}
\def\eqnr{\end{eqnarray}\noindent}
\def\beqr{\begin{eqnarray}}
 \def\beq{\begin{equation}}

\begin{document}

\title{Universal occurrence of mixed-synchronization in counter-rotating nonlinear coupled oscillators}
\author{Awadhesh Prasad\footnote{Email: awadhesh@physics.du.ac.in; 
Tel.: 91-11-27662752; Fax: 91-11-27667061}}
\affiliation{ Department  of Physics  and Astrophysics,
University of Delhi, Delhi 110007, India}
\begin{abstract} 
By coupling counter--rotating coupled nonlinear oscillators, we observe 
a ``mixed''  synchronization between the different dynamical variables of 
the same system. The phenomenon of amplitude death is 
also observed.  Results for coupled systems with co--rotating coupled oscillators 
are also presented for a detailed comparison. Results for  Landau-Stuart and \ro  oscillators are presented.
\end{abstract}
\maketitle

Natural systems are rarely isolated, and thus studies of coupled dynamical systems
which arise in a variety of contexts in the physical, biological, and social sciences,
have broad relevance to many areas of research.

\begin{figure}
\scalebox{0.4}{\includegraphics{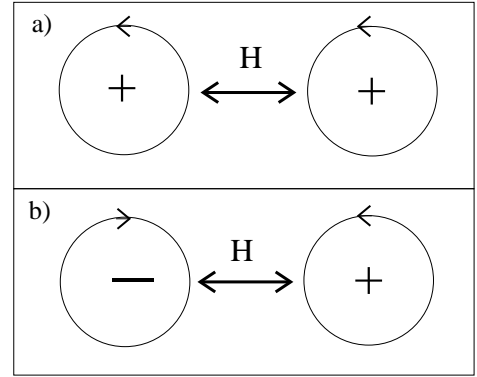}}
\caption{ Schematic coupling schemes for two coupled oscillators: (a) co-rotating
 and (b) counter-rotating. Arrows show the directions of rotation while $-$ and $+$  indicate the 
corresponding clockwise and anticlockwise rotations respectively.
 H denotes the coupling function, Eq. (\ref{eq:model}).}
\label{fig:sch}
\end{figure}

Consider the coupling of two nonlinear oscillators, schematically shown in Fig. \ref{fig:sch}, with the 
symbols `+' and `--' indicating the sense of rotation  of the orbits in the phase space, namely clockwise or counter-clockwise.  If directions of rotation of both oscillators are the same, as in
Fig. \ref{fig:sch}(a), the system is {\sl co}-rotating:  such coupled nonlinear oscillators 
have been extensively studied from both theoretical and experimental points of view \cite{synch}.

Important phenomena which have been observed for
co-rotating oscillators include synchronization, hysteresis,
phase locking, phase shifting, phase-flip, or riddling  \cite{synch,kaneko,ott,cs,hys,ap}. 
There are many forms of synchrony \eg complete synchronization: 
system variables become identical \cite{cs}, phase synchronization: phase difference 
between the oscillators becomes bounded \cite{ps}, and others \eg lag synchronization  
and generalized synchronization \etc\cite{synch}. In all these type of synchronizations if 
one of the dynamical variable is synchronized  then rest of the variables follow the same.

Another important property of co-rotating oscillators
is amplitude death (AD) \cite{bar} where  amplitude of the oscillation ceases to zero at 
stable fixed point of the system. This AD can be achieved with various type of
interactions between the oscillators \eg  mismatched oscillators \cite{ins}, 
delayed \cite{reddy}, conjugate \cite{rajat} and nonlinear \cite{mukesh} \etc

When the  rotations of the individual oscillators are  in opposite senses, as in 
Fig. \ref{fig:sch}(b), then the system is termed {\sl counter}-rotating.  To the best of our 
knowledge, counter-rotating systems  have  not explicitly been studied earlier, 
and in this Letter, we focus on such systems. 
We observe that both AD as well as synchronization arise with this form of coupling.
A novel ``mixed'' type of synchronization among different variables is seen to occur: 
some variables are synchronized in-phase while other variables can be out-of-phase.

Consider a coupled system of $N$ nonlinear oscillators specified by the equations  
\begin{eqnarray}
\dot{\mathbf{X}}_i=\mathbf{F}_i(\mathbf{X}_i,\omega_i)+\frac{\epsilon}{K_i} \sum_{j = 1}^{K_i}
A_{ij} \mathbf{H}(\mathbf{X}_i, \mathbf{X}_j),~~i=1,\ldots, N
\label{eq:model}
\end{eqnarray}
\noindent
where $\mathbf{X}_i$ is $m$-dimensional vector of dynamical variables and 
the $\mathbf{F}_i$'s specify the evolution equations of the different oscillators with
internal frequency parameter $\omega_i$.  Here $K_i$ is the number of connections to the
 $i-$th oscillators, namely its degree such that $1 \le K_i < N$, and $\epsilon$ is the coupling strength.
The connection topology is specified by the adjacency matrix ${\bf A}$, with  $A_{ij} = 1$ 
if oscillators $i$ and $j$ are
 connected to each other and $A_{ij} = 0$ otherwise.  The coupling function 
$\mathbf{H}: \R^m\to \R^m$ specifies the manner in which 
the oscillators $i$ and $j$ are coupled, with $\mathbf{H}(\mathbf{X}_i, \mathbf{X}_j)$ being a
 function  of $\mathbf{X}_i$ and $ \mathbf{X}_j$. 

\begin{figure}
\scalebox{0.45}{\includegraphics{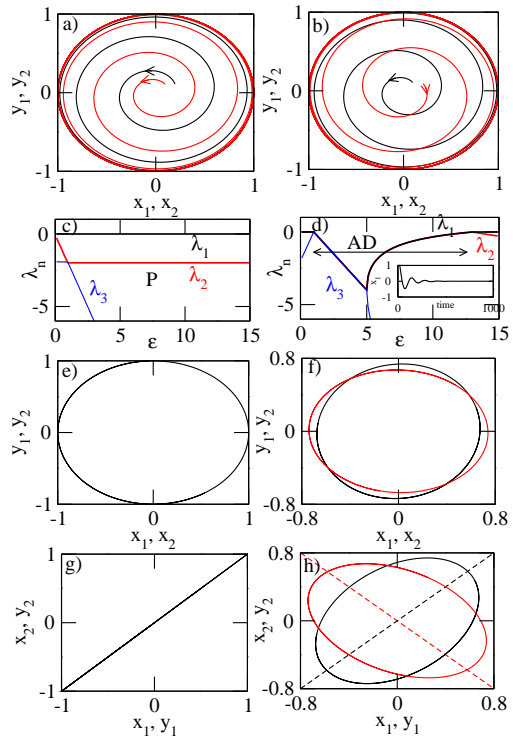}}
\caption{ Left and right panels correspond to the co- and counter-rotating respectively
 Landau-Stuart oscillators, Eq. (\ref{eq:ls}).
(a,b) The transient trajectories of uncoupled oscillators.
(c,d) A few largest Lyapunov exponents with coupling strength, $\ep$. Inset figure in (d)
 corresponds to the transient trajectories in AD regime at $\ep=2$.
(e,f) The asymptotic attractors in  $x_i-y_i$ plane. (g,h) The signature of
relative phase in $x_1$ vs $x_2$ and $y_1$ vs $y_2$ at $\ep=0.5$. 
 Dashed lines in (h) show the relative phase-difference of $0$ (black) and $\pi$ (red).}
\label{fig:ls}
\end{figure}

We first analyze two Landau--Stuart limit cycle oscillators \cite{ap}
coupled in the  co-- (Fig. \ref{fig:sch}(a)) as well as
counter-rotating (Fig. \ref{fig:sch}(b)) schemes. The coupled  equations are
\begin{eqnarray}
\dot{Z}_{i}&=&(1+i\omega_i-|Z_i|^2)Z_i+\ep (Z_j-Z_i)
\label{eq:ls}
\end{eqnarray}
\noindent
where $i,j=1,2$, $i\ne j$ and  $Z_i$ is the complex variables with real part
$x_i$ and imaginary part $y_i$.
In  Eq. (\ref{eq:model}) the variables are $X_i=[x_i,y_i]^T$ and
 $H=[(x_j-x_i), (y_j-y_i)]^T$ where $T$ denotes the transpose.
In the uncoupled case, \ie $\ep=0$, each oscillator has a fixed point at $Z_{*j}=0$ 
which is unstable with eigenvalues $1\pm i\omega_j$.
Irrespective of the values of $\omega_j$ each oscillator has the same attractor with
unit radius, and corresponding frequency $\omega_j$.
Shown in Fig. \ref{fig:ls} (a) and (b) are the transient trajectories for 
co-- ($\omega_1=\omega_2=5$) and  counter-rotating ($\omega_1=5, \omega_2=-5$)
respectively for two uncoupled oscillators. These clearly show that,
in these two cases, the trajectories are moving in the same  or in the opposite directions  respectively. 
In order to see the effects of coupling on these both schemes, results for co-- and counter 
rotating coupled oscillators are shown in the  
left and the right panels respectively in Fig. \ref{fig:ls}.  As the coupling is introduced, the
dynamical behavior changes. The few largest  Lyapunov exponents (see Fig. \ref{fig:ls}(c,d))
clearly show that in the case of co-rotating oscillators,
 the largest Lyapunov exponent remains zero for all coupling strengths
 which suggests the continuation of periodic motion.  This is shown Fig. \ref{fig:ls}(e) at $\ep=0.5$.
 However for counter-rotating oscillators, the maximal Lyapunov exponent 
(Fig. \ref{fig:ls}(d)) becomes negative (this is an instance of amplitude death, marked ``AD'') where the
fixed  point  $Z_{*1}=Z_{*2}=0$ is stabilized.  A simple eigenvalue ($\lambda$) calculation,  
with real part $\alpha$ and imaginary part $\beta$, gives  the relation, 
\begin{eqnarray}
\alpha=1-\ep\pm\ep \mbox{~~and~~}
\beta=\omega
\label{eq:ev}
\end{eqnarray}
\noindent
where $\omega_1=\omega_2=\omega$ for co-rotation (left panel).
For co-rotating coupled oscillators the real part of the
 largest eigenvalue,  $\alpha=1$, and hence fixed point will be unstable for
all values of $\omega$ and $\ep$. This rules out the possibility of AD in such scheme (for
details see Ref. \cite{ins}).  However  for $\omega_1=\omega, \omega_2=-\omega$ for the case 
of counter--rotating  oscillators (right panel) the eigenvalues will be
\begin{eqnarray}
\alpha=1-\ep \mbox{~~and~~} \beta= \sqrt{\omega^2-\ep^2} 
\label{eq:ev1}
\end{eqnarray}
\noindent
for $\ep <\omega^2$, while
\begin{eqnarray}
\alpha=1-\ep\pm\sqrt{\ep^2-\omega^2} \mbox{~~and~~}
\beta= 0,
\label{eq:ev2}
\end{eqnarray}
\noindent
for $\ep^2> \omega^2$. These relations give the region $1 < \ep< (1+\omega^2)/2$  where
the fixed point at the origin will be stable ($ \alpha < 0$)  as shown in
Fig. \ref{fig:ls}(d) by the arrow.
The transient trajectory in AD region at $\ep=2$ is
shown in the inset figure. This clearly
shows, which is a new result,  that AD is possible (via a Hopf bifurcation)
in the counter-rotating identical oscillators
while it is impossible in the co-rotating case. It should be noted that AD is possible
in co-rotating case only when there are mismatch in frequencies of the oscillators \cite{ins}.
Here we show that in counter-rotating case everything
is same as that of the co-rotating except directions of rotation. 
This novel result clearly distinguishes counter-rotating oscillators from 
the more extensively studied co-rotating case.

Another important result pertains to the regime of synchronized motion.
Figure \ref{fig:ls}(e) shows the attractors which are exactly same as uncoupled case 
(Fig. \ref{fig:ls}(a)) having complete synchronization. The relative phase between the
oscillators is shown in Fig. \ref{fig:ls}(g) where trajectories lie on the synchronization manifold, 
$x_1 = x_2, y_1= y_2$; this confirms the occurrence of complete synchronization.
However in the counter-rotating case,  Fig. \ref{fig:ls}(f) the attractors are  not identical: 
they do not remain as limit cycle circles (with unit radius)
instead  change shape as well as  get reduced in amplitude.  The oscillators are synchronized though,
as can be seen in the relative phase shown in Fig. \ref{fig:ls}(h). The $x-$ variables
essentially in-phase while $y-$ variables are out-of-phase (dashed lines  correspond to the 
relative phase  zero (black) and $\pi$ (red)). The simultaneous  presence of nearly 
in- as well as out-of-phase synchrony in different variables among the same oscillators
 is termed here as ``mixed" synchronization.
 
 It is important to note that the usual definition of phase difference between oscillators
is not applicable  here since there is continuous change of individual angle in opposite directions.
Therefore the effect of counter-rotation has drastic difference due to presence of
mixed-synchronization as compared to co-rotation.

\begin{figure}
\scalebox{0.4}{\includegraphics{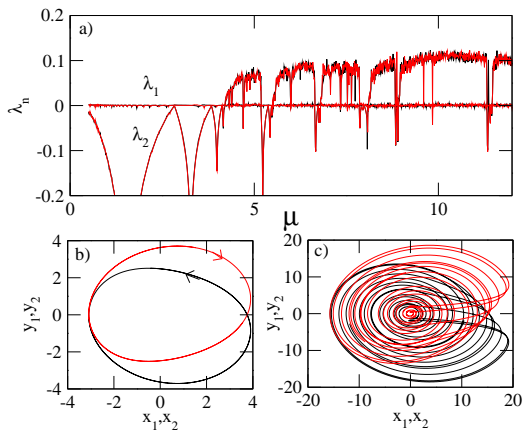}}
\caption{ (a) The largest two Lyapunov exponents with internal parameter, $\mu,$
 for anticlockwise ($\omega_1=1$ -- black lines) and clockwise ($\omega_2=-1$ -- red lines) 
for uncoupled ($\epsilon=0$) \ro oscillators, Eq. (\ref{eq:ross}).  The trajectories of 
individual oscillators for (b) periodic ($\mu=2$) and (c) chaotic ($\mu=10$) motions.}
\label{fig:ross-single}
\end{figure}

The generality of these results is demonstrated in a system of coupled \ro oscillators \cite{ross},
given by the equations of motion
\begin{eqnarray}
\nonumber
 d x_i(t)\over dt &=& -\omega_i y_i -z_i+ \epsilon(x_j-x_i)\\ \nonumber
 d y_i(t) \over dt &=& \omega_i x_i +0.2 y_i\\
d z_i(t)\over dt &=& 0.2+ z_i (x_i - \mu)
\label{eq:ross}
\end{eqnarray}
where $\omega_i$ (internal frequency) and $\mu$ are parameters, and $i,j=1,2$, $i\ne j$. 
(With regard to Eq. (\ref{eq:model}), $X_i=[x_i,y_i,z_i]^T$ and $H=[(x_j-x_i),0,0]^T$.)

Changing the parameter $\omega_i$  merely changes the sense of rotation; as can be seen 
in Fig. \ref{fig:ross-single}(a)  the two largest Lyapunov exponents as a function of  $\mu$ are essentially
identical for   $\omega_1=1$ (black curves) and $\omega_2=-1$ (red curves). 
Corresponding typical trajectories for  periodic and chaotic  motions are shown in 
Fig. \ref{fig:ross-single}(b) and (c) at $\mu=2$ and $10$ respectively. 
Note that there is change of position of the attractors in the phase space. These 
fixed points $(x_{*i}=(\mu\pm\sqrt{\mu^2-0.16/{\omega_i^2}})/2, 
y_{*i}=5\omega_i x_{*i}, z_{*i}= 5 \omega_i^2x_{*i})$  depend on the
sign of $\omega_i$ (unlike the case of Eq. (\ref{eq:ls}) where the origin fixed point
is independent of frequency).

\begin{figure}
\scalebox{0.45}{\includegraphics{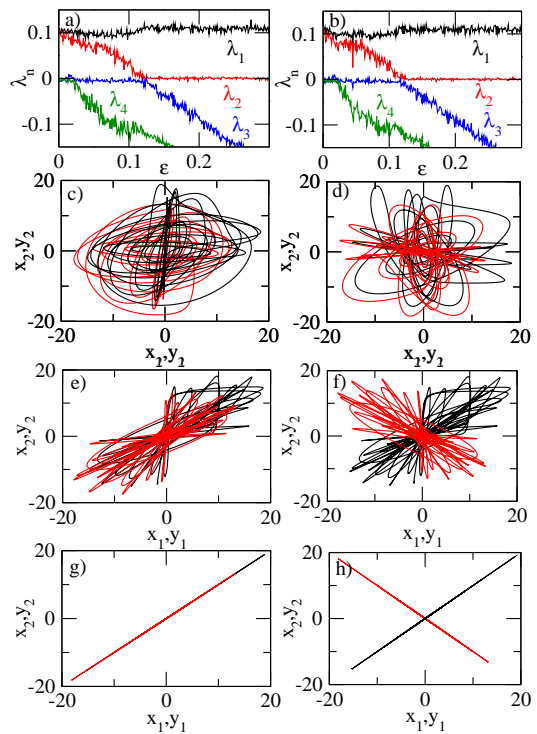}}
\caption{ Left and right panels correspond to the co- and counter-rotating chaotic \ro 
oscillators respectively, Eq. (\ref{eq:ross}).
(a,b) A few largest  Lyapunov exponents with coupling strength, $\epsilon$ at $\mu=10$.
The signature of (c,d) non-synchronization, (e) phase-synchronization, (f) mixed-synchronization,
(g) complete-synchronization and (h) mixed-synchronization. The  coupling strengths for
(c,d), (e,f) and (g,h) are at $\ep=0.005, 0.05$ and $0.2$ respectively.}
\label{fig:ross-coupled}
\end{figure}

The change in dynamical behavior arising from the coupling is shown
 in Fig. \ref{fig:ross-coupled} for  co-- (left panel) 
and counter--rotating (right panel) coupled chaotic ($\mu=10$) \ro
 oscillators, Eq. (\ref{eq:ross}).   For small coupling, say $\ep=0.005$, the
motion is unsynchronized which can be seen from the relative phase between
the trajectories in  Fig. \ref{fig:ross-coupled}(c)  in $x_1$ vs $x_2$ (black) and 
$y_1$ vs $y_2$ (red).  As the coupling strength is increased
beyond $\ep \sim 0.02$ phase-synchronization sets in (the fourth largest Lyapunov exponent (in green) 
becomes negative \cite{ps}), and after this all variables $x,y,z$  remain in--phase 
as shown in  Fig. \ref{fig:ross-coupled}(e). When the coupling is increased beyond 
$\ep \sim 0.12$ complete synchronization occurs \cite{ps} as seen in 
Fig. \ref{fig:ross-coupled}  (g) where $x_1=x_2,y_1=y_2$.

For counter-rotating  oscillators ($\omega_1=1, \omega_2=-1$) (right panel) 
 Lyapunov spectrum (Fig. \ref{fig:ross-coupled}(b)) remains the same as for
co-rotating oscillators (Fig. \ref{fig:ross-coupled}(a)) showing  that  the
invariant measure does not change.  For small coupling the oscillators are not in the phase 
synchronized state Fig. \ref{fig:ross-coupled}(d),  as analogous to the  co-rotating
case  Fig. \ref{fig:ross-coupled}(c).  However,  
 as  coupling is increased further, the  phase synchronization sets in,
 (Fig. \ref{fig:ross-coupled}(f)), but the relative phase between the three variables is not identical.

\begin{figure}
\scalebox{0.4}{\includegraphics{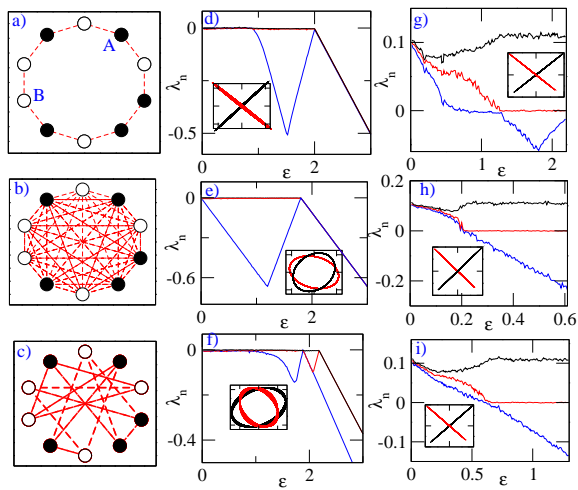}}
\caption{ Left panel shows the network of different topologies for $N=10$ coupled oscillators:
 (a) periodic chain (b) globally coupled and (c) small-world network--see
 the text for details. The largest three Lyapunov exponents with coupling strength
 for different topologies are plotted in the middle  and right panel corresponding to 
the Landau-Stuart  and  the \ro coupled oscillators respectively.  
Inset figures show the relative phase of  $x_A$ vs $x_B$ (black) and $y_A$ vs $y_B$ (red) between 
the marked oscillators A and B.
The coupling strengths for the inset figures of middle  panels (d,e,f) are at $\ep=1.5$ while
the corresponding values for  (g), (h) and (i) are at $\ep=2, 0.5$ and $1$ respectively.}
\label{fig:mix}
\end{figure}

Shown in Fig. \ref{fig:ross-coupled} (f) is the relative phase in $x_1-x_2$ 
and $y_1-y_2$ planes.  Clearly $x_1$ and $x_2$ are in--phase (black) 
synchronized while $y_1$ and $y_2$ are out-of-phase (red).  This is clearer for higher
 coupling,  $\ep=0.2$ (Fig. \ref{fig:ross-coupled}(h)) where there is
 complete synchronization in $x_1$ and $x_2$ (zero relative phase) while
$y_1$ and $y_2$ remain out-of-phase with phase difference  $\pi$.
We find $z_1$ and $z_2$ in complete synchrony (results not shown here \cite{new}).
For counter-rotating coupled oscillators it appears that if one variable is
 in-phase synchrony, then at least one other variable must be out-of-phase,
\ie if motion is on the manifold ($x_1=x_2, y_1=y_2, z_1=z_2$) for co-rotating case 
 (Fig. \ref{fig:ross-coupled}(g)) then the corresponding manifold for the  counter-rotating case
is ($x_1=x_2, y_1=-y_2, z_1=z_2$), Fig. \ref{fig:ross-coupled}(h).

Mixed-synchronization and AD also occur in network of oscillators where the rotation sense is
chosen randomly. We present representative results for  $N=10$  for Landau-Stuart 
and \ro oscillators in middle and right panels respectively in
Fig. \ref{fig:mix}. The rotations of individual oscillators are indicated in
 Fig. \ref{fig:mix}(a,b,c) where open and filled circles correspond to the
clockwise and anticlockwise rotations respectively in these particular realizations.
Three different topologies are taken: (i) nearest neighbor coupling with periodic boundary 
conditions ($K=2$)--Fig. \ref{fig:mix}(a); (ii) global coupling where each oscillator is coupled
with all  others $(K=N-1)$ -- see Fig. \ref{fig:mix}(b); (iii) each oscillator is   coupled to  three
 ($K=3$) random nodes --- see Fig. \ref{fig:mix}(c) which realizes a small world 
 topology \cite{sw}. The coupling functions $H$ for these 
network of oscillators are the same as in Figs. \ref{fig:ls} and \ref{fig:ross-coupled} for 
Landau-Stuart and \ro oscillators respectively. 
In Figs. \ref{fig:mix}(d) to (i) the largest few Lyapunov exponents are plotted as a function of
the coupling strength in the three cases.
The inset figures in Fig. \ref{fig:mix}(d-i)  show the relative phase $x_A$ vs $x_B$ (black curves)
 and $y_A$ vs $y_B$ (red curves)  for  counter-rotating oscillators between
the marked oscillators A and B in (a) verifying the existence of  mixed-synchronization for both
systems.  The occurrence of AD is also observed in networks of
Landau-Stuart oscillators, as shown by negative Lyapunov exponents in the middle panel. 

In summary, we have observed the universal  occurrence of mixed-synchronization
in counter-rotating oscillators: some variables are synchronized in-phase, while others
are out-of-phase.  The phenomenon of amplitude death is also observed in some cases,
as for example, the Landau-Stuart limit cycle oscillators studied here. These results are 
 also found to persist in networks of oscillators of different topologies.
We have also checked that these results persist under parameter mismatch  \cite{new},
and believe that they are generally applicable and should be observable in experiments. 
Thus the phenomena observed here will occur in other  coupled systems as well.

In this paper we have considered only linear diffusive coupling, and other types of interactions
such as one-way, conjugate \cite{rajat},  and nonlinear couplings \cite{mukesh} are still need to be explored.
Dynamical phenomena  such as  riddling, hysteresis, phase locking, phase-shifting, phase-flip  \cite{synch,kaneko,ott,cs,hys,ap} \etc which have been seen in co-rotating coupled systems 
should also be explored in counter-rotating coupled oscillators.
 
This work is supported by the Delhi  University and the Department of Science and 
Technology, Government of India. I thank R. Ramaswamy for his 
comments on this paper.

\end{document}